\documentclass[onecolumn]{aastex631}

\usepackage{multirow}
\usepackage[flushleft]{threeparttable}
\usepackage{graphicx}	
\usepackage{amsmath}	
\usepackage{amssymb}	

\newcommand{\batman}{\textsc{batman}}
\newcommand{\hdb}{HD\,189733\,b}

\shorttitle{X-ray transit profiles}
\shortauthors{King et al.}

\begin{document}

\title{Generating X-ray transit profiles with \textsc{batman}}

\author[0000-0002-3641-6636]{George W. King}
\affiliation{Department of Astronomy, University of Michigan, Ann Arbor, MI 48109, USA}
\affiliation{Department of Physics, University of Warwick, Gibbet Hill Road, Coventry, CV4 7AL, UK}
\affiliation{Centre for Exoplanets and Habitability, University of Warwick, Gibbet Hill Road, Coventry, CV4 7AL, UK}

\author{L\'{i}a R. Corrales}
\affiliation{Department of Astronomy, University of Michigan, Ann Arbor, MI 48109, USA}

\author[0000-0003-1452-2240]{Peter J. Wheatley}
\affiliation{Department of Physics, University of Warwick, Gibbet Hill Road, Coventry, CV4 7AL, UK}
\affiliation{Centre for Exoplanets and Habitability, University of Warwick, Gibbet Hill Road, Coventry, CV4 7AL, UK}

\author[0009-0009-0960-6280]{Raven C. Cilley}
\affiliation{Department of Astronomy, University of Michigan, Ann Arbor, MI 48109, USA}

\author{Mark Hollands}
\affiliation{Department of Physics, University of Warwick, Gibbet Hill Road, Coventry, CV4 7AL, UK}



\begin{abstract}

We present an adaptation of the exoplanet transit model code \textsc{batman}, in order to permit the generation of X-ray transits. Our underlying extended coronal model assumes an isothermal plasma that is radially symmetric. While this ignores the effect of bright, active regions, observations of transits in X-rays will require averaging across multiple epochs of data for the foreseeable future, significantly reducing the importance of more complex modelling. Our publicly available code successfully generates the predicted W-shaped transit profile in X-rays due to the optically thin nature of the emission, which concentrates the expected observational emission around the limb of the photospheric stellar disc. We provide some examples based on the best known X-ray transit target, \hdb, and examine the effect of varying the planet size, coronal temperature, and impact parameter on the resulting transit profile. We also derived scaling relationships for how the overall transit detectability is affected by changing these parameters. Over most of the parameter space, we find that the detectability scales linearly with the cross-sectional area of the planet in X-rays. The relationship with increasing coronal temperature is less fixed, but averages out to a power law with slope $-1/4$ except when the impact parameter is high. Indeed, varying impact parameter has little effect on detectability at all until it approaches unity.


\end{abstract}




\section{Introduction}
\label{sec:intro}

Exoplanets have been observed transiting their stars across a wide array of wavelengths over the past couple of decades, with the aim of investigating the nature of their atmospheres. Studies have primarily focussed on visible and near-infrared wavelengths, but a small handful of systems have been observed at shorter wavelengths, such as the near-ultraviolet \citep[e.g.][]{Fossati2010,Sing2019,Wakeford2020,189OM} and far-ultraviolet \citep[e.g.][]{Vidal-Madjar2003,Linsky2010,Ehrenreich2015,Bourrier2018}. Some of these more underexplored wavelengths have the power to probe extended or escaping atmospheres as evidenced by e.g. deeper transits than the optical, and asymmetric transit profiles.

X-rays, together with extreme-ultraviolet (EUV; referred to together as the XUV), can drive the escape of material in the upper atmospheres of exoplanets, especially those orbiting within a few tenths of an AU of their host star \citep[e.g.][]{Lammer2003,Owen2012}. Escaping material has been observed through deeper transits in a handful of specific wavelengths such as Ly-$\alpha$, where transits as deep as 56\% have been observed \citep{Ehrenreich2015}. Asymmetries have also been observed, and attributed to large comet-like tails of escaped material \citep[e.g.][]{Lavie2017,Owen2023}. Transits at the escape-driving XUV wavelengths have however proved difficult to detect, primarily due to the typically low fluxes of their FGKM type stars. \citet{Poppenhaeger2013} presented evidence of a drop in the X-ray count rate of the best known X-ray transit candidate \citep{Cilley2024}, HD\,189733, during the time of primary transit, by combining six \textit{Chandra} observations and one \textit{XMM-Newton} observation. However, even the combined light curves were noisy, and consequently neither the transit shape nor the depth were well constrained.

In theory, one expects a double dip, or W-shaped, transit profile due to a {\em limb brightening} effect. The two minima in the transit profile correspond to the points at which the planet passes over the edge of photospheric disk. These regions are brightest because the coronal plasma which emits in X-rays is optically thin, and immediately outside the photospheric disk is where one observes through the greatest column density of coronal plasma. 

Early models of transits across a limb brightened disk focused on chromospheric lines, with the emission modeled as a geometrically thin ring, assumed not to extend beyond the stellar disk \citep{Assef2009,Schlawin2010}. For X-rays, this assumption is not necessarily appropriate as the corona extends out above the photosphere. A handful of more studies since have more specifically looked at the properties of X-ray transit profiles. \citet{Llama2015} simulate transits based on the resolved X-ray properties of the Sun, including investigating the effects of a planet crossing active and inactive regions of the corona. \citet{Marin2017} model the X-ray transit of prototypical transiting hot Jupiter \hdb, using an isothermal coronal model extending beyond the photospheric disk, as we will also adopt in this work.

\textsc{batman} \citep{batman} is a Python package for generating exoplanet transit light curves\footnote{See also \url{https://lkreidberg.github.io/batman/docs/html/index.html}.}, which has seen wide implementation in the literature. 
The \textsc{batman} code assesses the coverage of the stellar disk by the planet by calculating the covering fraction in each of a series of concentric rings, and any radially symmetric limb darkening law can be implemented within its algorithm.

We adapt \textsc{batman} in order to simulate X-ray transits, by using its custom ``limb darkening" law to instead implement a simple, radially symmetric stellar coronal emission model. This work enables the relatively fast generation of X-ray light curves based on coronal and planet properties, and our code, including our custom implementation of \textsc{batman}, is now publicly available\footnote{\url{https://github.com/georgewking/batmanX-rays}}. Following a description of our model and calculations, we give some examples of using the code to generate transit models of \hdb, and describe some of the limitations of the model and possible ways to mitigate them. We also refer the reader to a companion paper by \citet{Cilley2024}, which uses the model described here to generate simulated data for hundreds of known exoplanet systems, identifies which are the best targets for observation, and demonstrates statistically the effect of varying the model parameters on the detectability of the transits.

\section{Model}
\label{sec:model}

\begin{figure}
\centering
 \includegraphics[width=0.68\columnwidth]{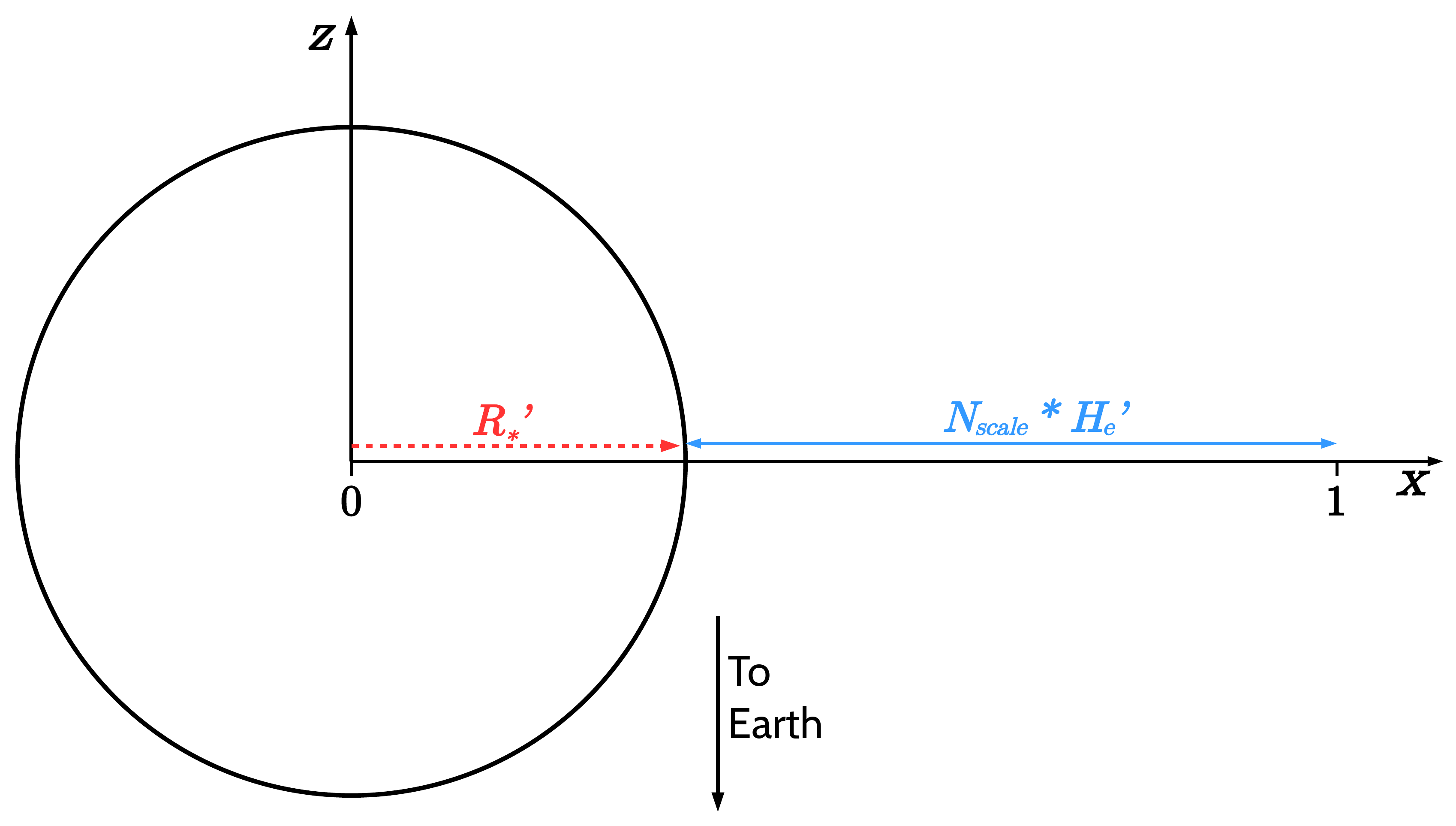}
  \caption{The considered geometry of the coronal profile in the $0 < x < 1$ coordinate system used by \textsc{batman}. $N_{\rm scale}$ is the number of emission scale heights the model is calculated out to, which is user defined. We use $N_{\rm scale} = 6$ throughout the examples shown in this work. The stellar radius in this coordinate system, $R'_{\rm *}$, is defined as 1-($N_{\rm scale}$ * $H'_{\rm e}$).}
 \label{fig:batmanGeo}
\end{figure}

We modelled the stellar corona with an exponentially decaying density profile, with 3D radial symmetry. The resulting coronal emission flux, $f_\text{X}$, is described as
\begin{equation}
f_\text{X}(r) = \begin{cases}
f_\text{X}(0) \exp(\frac{-2r}{H_{\rm p}}), &\text{for } r > R_*.\\
0, &\text{for } r < R_*
\end{cases}
\label{eq:coronaProfile}
\end{equation}
where $r$ is the radial distance from the stellar centre,  $R_*$ is the stellar radius (to the photosphere edge), and  $H_{\rm p}$ is the pressure scale height of the corona. We emphasize this is a radial function in 3D, and does not describe the projection of the emission in 2D. On the sky, a significant contribution to the overall emission can come from within the projection of the photospheric disk. The factor of two in the exponent comes from $f_\text{X}$ being proportional to $n^2_e$, the square of the election number density. Assuming an isothermal plasma in hydrostatic equilibrium, $H_{\rm p}$ is equal to the electronic density scale height \citep[e.g.][]{Marin2017}. Under the isothermal assumption, this is given by
\begin{equation}
    H_{\rm p} = \frac{2 k_B T}{\mu m_H g},
\label{eq:scaleH}
\end{equation}
where $k_B$ is the Boltzmann constant, T is the temperature of the coronal plasma, $\mu$ is the mean particle weight, $m_H$ is the mass of the hydrogen atom, and $g$ is the surface gravity of the star, which we assume is constant \citep[as in e.g.][]{Marin2017}. Hydrostatic equilibrium and constant gravity are reasonable first order assumptions for the lowest part of the corona where the majority of the quiescent emission emanates from. Complex coronal models that include non-hydrostatic equilibrium, active regions and flares are beyond the scope of this work.
Through the rest of the paper, we prefer to use the emission scale height, $H_{\rm e}$, as the emission is the property we are directly modeling. $H_{\rm e}$ is given by $H_{\rm p}/2$, as per the factor of two in Equation \ref{eq:coronaProfile}. 

In its standard use, \batman\ uses a 2D radially symmetric coordinate system defined for $0 < x < 1$, where $x=0$ corresponds to the centre of the stellar disc, and $x=1$ corresponds to the edge of the photospheric disc. Given that coronal emission originates external to the photospheric disc, we adjusted $x=1$ in our model to be a user-defined number, $N_{\rm scale}$, of emission scale heights from the edge of the photosphere along the plane of the sky. The geometry we used is displayed in Fig. \ref{fig:batmanGeo}. The stellar radius, and thus edge of the photospheric disk, in this coordinate system, $R'_{\rm *}$, is therefore
\begin{equation}
    R'_{\rm *} = 1-(N_{\rm scale} * H'_{\rm e}),
\end{equation}
where $H'_{\rm e}$ is the emission scale height in this coordinate system. This can be calculated from physical, stellar radii units using
\begin{equation}
    H'_{\rm e} = \frac{H_{\rm e}}{1 + (N_{\rm scale} * H_{\rm e})}
\end{equation}

In order to project our coronal profile into two dimensions, we integrated the coronal profile along the line of sight ($z$ in Fig. \ref{fig:batmanGeo}). Line of sight integrals were calculated at 10,001 equally spaced points in $x$ between 0 and 1, stretching out radially on the projection. The line of sight integrals are given by
\begin{equation}
I(x) = \begin{cases}
2I_0\int^{\infty}_0 \exp\left(-\frac{1}{H'_{\rm e}}\sqrt{z^2 + x^2}\right)\ dz, &\text{for } x > R'_{\rm *},\\
I_0\int^{\infty}_{\alpha} \exp\left(-\frac{1}{H'_{\rm e}}\sqrt{z^2 + x^2}\right)\ dz, &\text{for } x < R'_{\rm *},
\end{cases}
\label{eq:integral}
\end{equation}
where 
$I_0$ is a normalization factor, while $\alpha$ is given by
\begin{equation}
\alpha = \sqrt{R^{\prime\ 2}_* - x^2}.
\end{equation}
The solution to the integral in the top line of Equation \ref{eq:integral} is a modified Bessel function of the second kind. Specifically, for $x > R'_{\rm *}$
\begin{equation}
I(x) = 2 I_0 x K_1 \left(\frac{x}{H'_{\rm e}}\right).
\end{equation}
The expression in the second line of Equation \ref{eq:integral} does not integrate to a simple analytic equation, and so we evaluate it numerically. These values must be recalculated separately for each $H'_{\rm e}$. The resulting intensities were normalized such that the total emission intensity is unity, as per the requirements for \batman:
\begin{equation}
    \int^{2\pi}_0 \int^1_0 I(x)x\,dx\,d\theta = 1.
    \label{eq:normInt}
\end{equation}
An array of the 10,001 normalized integration results are passed as a custom limb darkening law in the \textsc{batman} code \citep{batman}, and some minor editing of the \_custom\_ld.c file was required in order to handle this change. For each $x$ examined by \textsc{batman} during a calculation, a linear interpolation of the surrounding integration values was performed to estimate the relative brightness of the corona at that $x$ for the current model being considered, as implemented in \_custom\_intensity.c file. 


\section{Example results with \hdb}

\begin{table}
  \caption{Adopted stellar and planetary parameters for HD\,189733(b).}
  \label{tab:189param}
  \centering
  \begin{threeparttable}
  \begin{tabular}{l l c l c}
    \hline
    Parameter & Symbol & Value & Unit & Ref. \\
    \hline
    Stellar mass & $M_*$            & $0.823\pm0.029$  & M$_\odot$ & 1\\
    Stellar radius & $R_*$ & $0.780^{+0.017}_{-0.024}$ & R$_\odot$ & 2\\
    Optical planet to star rad. & $R_{\rm p,opt}/R_*$ & $0.15641\pm0.00010$$^\dagger$ & & 3\\
    Orbital period & $P_{\rm orb}$ & 2.218575200(77) & d & 4\\
    Semi-maj. axis & \multirow{2}{*}{$a/R_*$} & \multirow{2}{*}{$8.863\pm0.020$} &  & \multirow{2}{*}{5}\\
    to star radius & & & & \\
    Eccentricity & $e$                         & 0  &               & 6\\
    Orbital inclin. & $i$              & $85.710\pm0.024$  & $^\circ$ & 5\\
    Impact parameter    & $b$ & $0.6636\pm0.0019$ &  & 4\\
    \hline
\end{tabular}
\begin{tablenotes}
\item References: (1) \citet{Triaud2009}; (2) \citet{GaiaDR2}; (3) \citet{Sing2011}; (4) \citet{Baluev2015}; (5) \citet{Agol2010}; (6) \citet{Bouchy2005}.
\end{tablenotes}
\end{threeparttable}
\end{table}

\begin{figure*}
    \centering
    \includegraphics[scale=0.473]{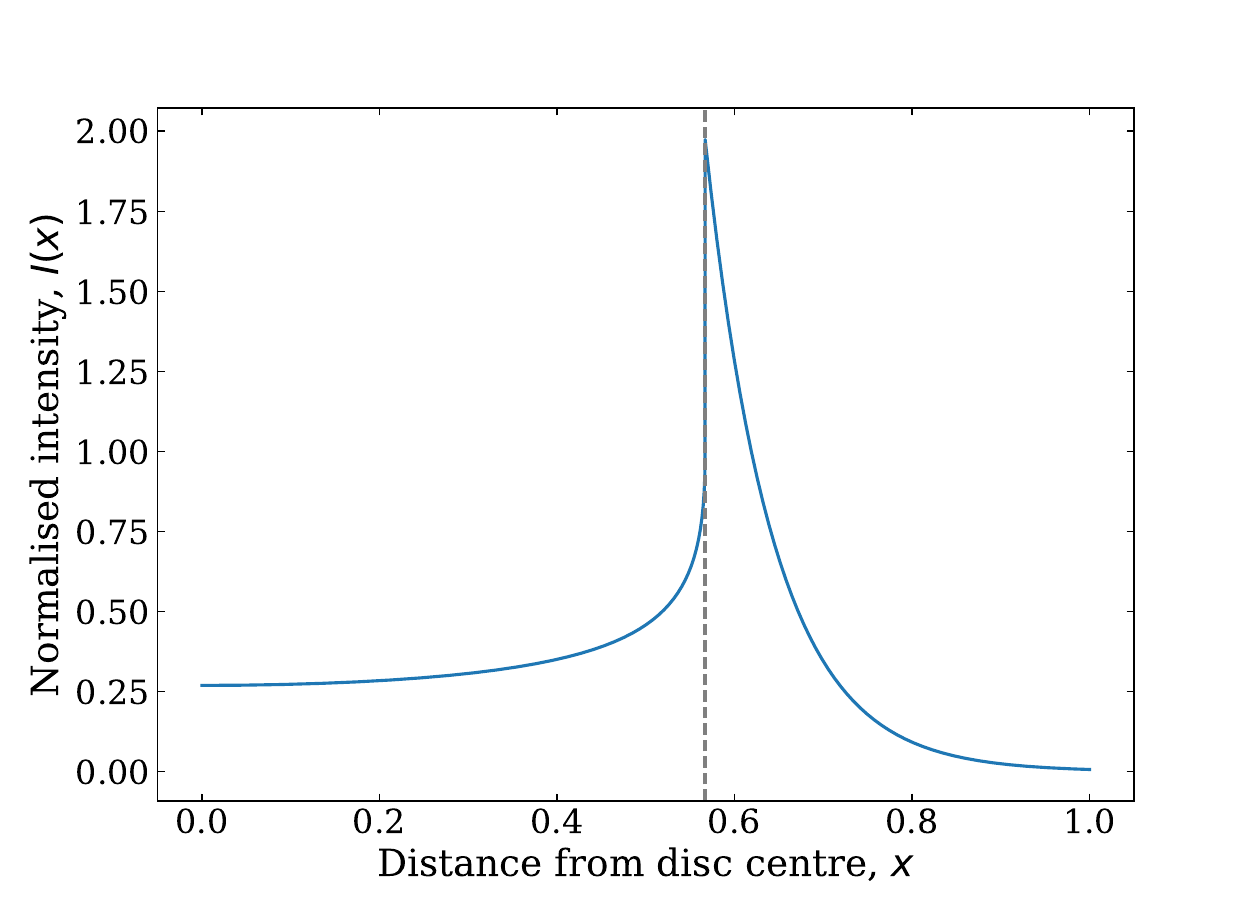}
    \includegraphics[trim={1.6cm 2.8cm 0cm 1.6cm},clip,scale=0.297]{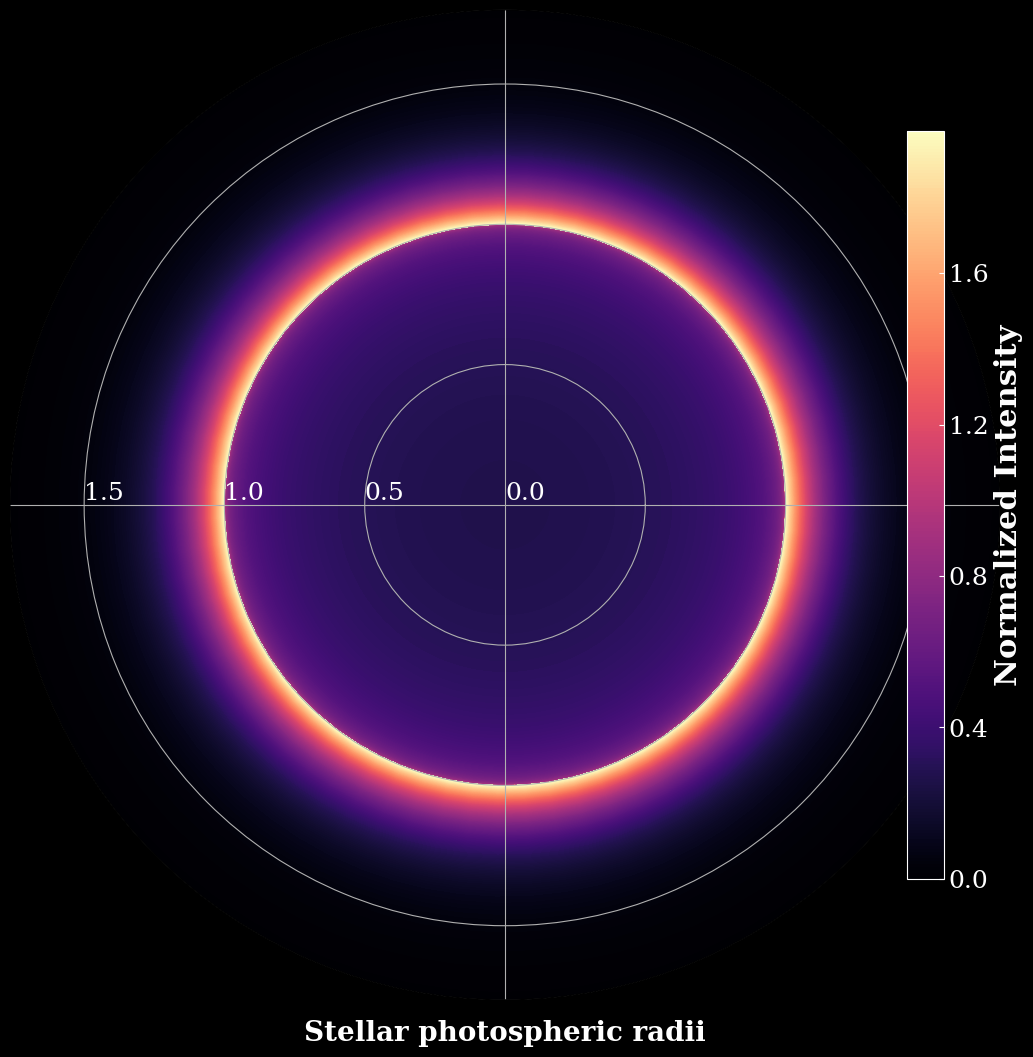}
    \caption{1D (left panel) and 2D (right panel) schematic of the normalized emission intensities for HD189733 for $kT=0.34$\,keV, $H_{\rm e} = 0.127$\,R$_*$. In the 1D plot, the distance from the stellar disc centre in is plotted in batman's 0 to 1 coordinate system, with the edge of the stellar disc displayed with the grey dashed line. In the 2D plot, the distance units are in stellar radii.}
    \label{fig:intensityPlots}
\end{figure*}

To demonstrate our model, we present a few examples of our X-ray transit framework for \hdb. 
The adopted parameters of the HD\,189733 system are given in Table \ref{tab:189param}. 
Except where otherwise specified, we adopt $H_{\rm e}$ = 0.127\,R$_*$, based on a temperature of $kT = 0.34$\,keV. This value is the mean of the three temperatures in the HD\,189733 X-ray spectral fit by \citet{Bourrier2020}, weighted by their respective Xspec normalization values, which are in turn directly related to their emission measures. In all examples, we calculated the coronal emission out to six emission scale heights ($N_{\rm scale} = 6$). Fig.~\ref{fig:intensityPlots} displays the 1D and 2D distributions of the coronal emission intensity for $H_{\rm e} = 0.127$\,R$_*$. The grey dashed line on the 1D plot is the edge of the stellar photosphere. The maximum in coronal intensity 
also occurs at this radial position.

This radial distribution of the coronal emission is used with the \textsc{batman} architecture, as described in Section~\ref{sec:model}. For each example light curve, we evaluated the coronal flux as an opaque disk crosses it, using 500 equally spaced points from 0.9 to 1.1 in orbital phase. For the HD\,189733 system, this corresponds to 1.3\,min between sample points.

\subsection{Fiducial case}
\label{sec:Fiducial}

\begin{figure*}
    \centering
    \includegraphics[scale=0.64]{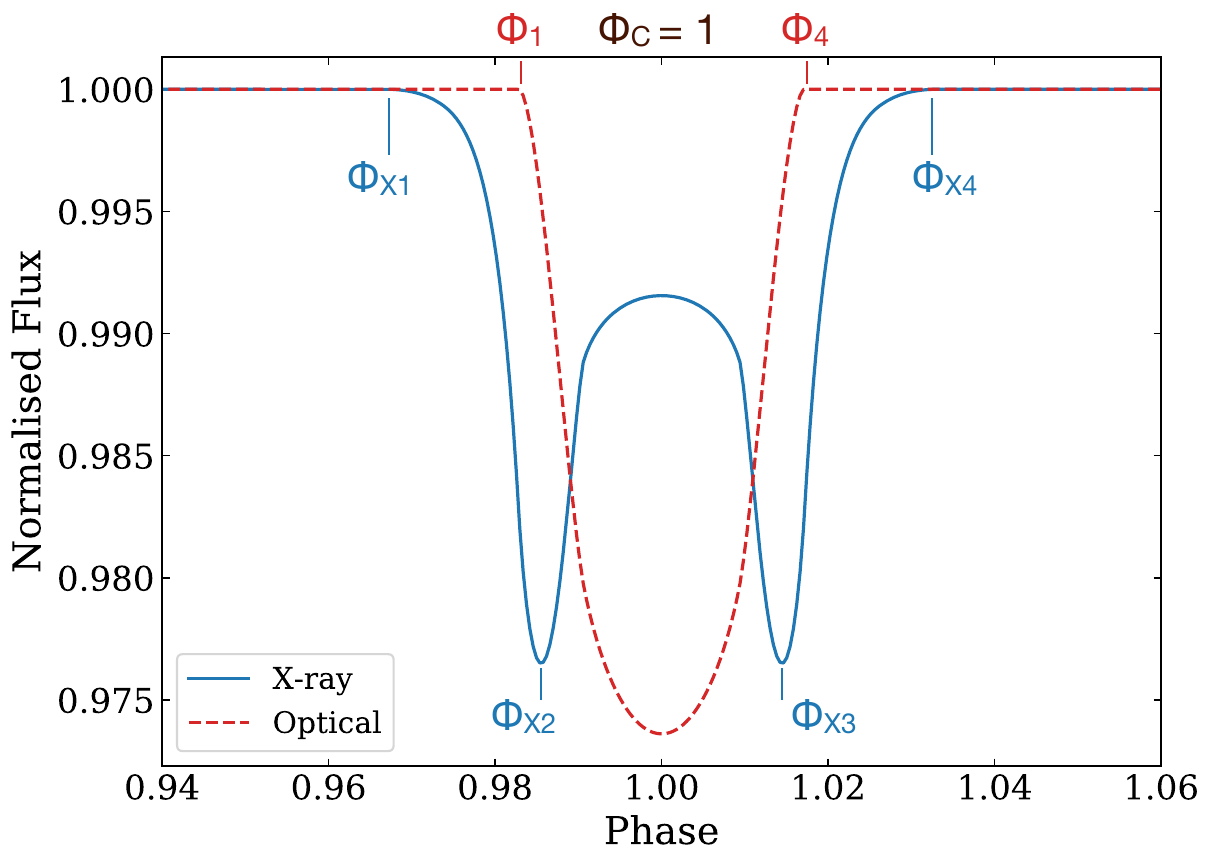}
    \caption{Comparison of optical and X-ray transit models for HD\,189733. The annotations mark key moments where the X-ray transit shape changes characteristics  (blue) due to the relative position of the planetary disk and stellar photosphere, analogous to key moments in the optical transit (red). In the case of \hdb, the impact parameter is high enough that optical transit light curve never goes flat, making it so the $\Phi_2$ and $\Phi_3$ points are not readily apparent. At both wavelengths, the transit midpoint ($\Phi_C$) are the same. 
    Calculated transit characteristics associated with these models are given in Table \ref{tab:combRes}.
    }
    \label{fig:optR}
\end{figure*}

First, we examine the case that the effective radius of the X-ray absorbing disc of \hdb, R$_{\rm p,X}$, is the same as the optical radius, R$_{\rm p,opt}$. 
Fig.~\ref{fig:optR} shows a comparison between X-ray and optical transit light curves for \hdb. For the optical limb darkening coefficients, we used the \citet{Claret2011} model for a star with $T_{\rm eff} = 4800$\,K, $\log g = 4.5$\,cm\,s$^{-2}$, and Solar metallicity, as observed with the Johnson V filter.

Fig.~\ref{fig:optR} also presents a timing framework for X-ray transits analogous to that used for optical transits. While the start of optical ingress (end of egress) is labelled $\Phi_1$ ($\Phi_4$) occurs at first contact between the projected image of the planetary disk and the stellar photosphere, the extended nature of the stellar corona makes it impossible to identify the exact beginning and end of the X-ray transit contact point. Based on our model and our choice of $N_{\rm scale}$, we instead define $\Phi_{X1}$ and $\Phi_{X4}$ as the points where the leading and trailing limbs of the X-ray absorbing region of the planet cross $6 H_{\rm e}$ above the photosphere for the first and last time, respectively. While non-zero, the emission outside of this distance in the projection should be a negligible contribution to the overall X-ray flux of the star. The $\Phi_{X2}$ and $\Phi_{X3}$ mark the times when the planetary disk is blocking the maximum amount of X-ray emitting coronal plasma, due to limb brightening. At these points, the center of the planetary disk is usually just outside the disc of the stellar photosphere in projection. When the planet traverses the regions inside the disc of the stellar photosphere, it blocks a smaller column of X-ray emitting plasma compared to the limbs, leading to the unusual `W' light curve shape that is characteristic of X-ray transits. The transit mid-point, $\Phi_C$, is expected to be the same in the X-ray as the optical, assuming spherical symmetry in the X-ray absorbing portion of the planetary atmosphere.

As we will demonstrate in Section \ref{sec:scalings}, the exact shape, timing, and depth of X-ray transits depend strongly on the properties of the stellar coronae and planetary absorber. For the purposes of comparison, we define the following properties as benchmarks. Code to calculate these quantities is included in our GitHub release.
The {\em full duration} of the transit is defined as:
\begin{equation}
    \Phi_D \equiv \Phi_{X4} - \Phi_{X1}
\end{equation}
in units of the orbital period. 
The average depth of the transit is thereby
\begin{equation}
    \bar{\delta} \equiv \frac{1}{\Phi_D} \int 1 - \frac{F(\phi)}{F_0(\phi)}\ d\phi. 
\end{equation}
We use the average transit depth to define a unit-less quantity, {\em flux blocked}, as
\begin{equation}
\label{eq:FluxBlocked}
    \chi \equiv \bar{\delta}\ \Phi_D
\end{equation}
which describes the fraction of stellar light blocked by the transiting planet over the course of one orbital period. Even though this small value is not one that would typically be measured, it provides us with a means to directly compare the detectability of the different models we examine. An estimate for the total transit signal (i.e., integrated flux missing over the course of the transit) is then:
\begin{equation}
    \Delta F_X = \chi P F_0
\end{equation}
where $P$ is the orbital period of the planet and $F_0$ is the predicted X-ray flux without the transit. We present the unit-less value, $\chi$, throughout this work so that an X-ray transit signal may be estimated for any planetary system of interest.

We define additional benchmark parameters that characterize the morphology of the transit light curve, which might also affect transit detectability. 
We define the mid-transit depth as
\begin{equation}
    \delta_{X,{\rm mid}} \equiv 1 - \frac{F(\Phi_C)}{F_0(\Phi_C)},
\end{equation}
and the maximum depth of the transit light curve as
\begin{equation}
    \delta_{X,{\rm max}} \equiv 1 - \frac{F(\Phi_{X2})}{F_0(\Phi_{X2})} \equiv 1 - \frac{F(\Phi_{X3})}{F_0(\Phi_{X3})}.
\end{equation}
In most cases, $\delta_{X,{\rm max}}$ occurs at $\Phi_{X2}$ and $\Phi_{X3}$, and we define the {\em trough-to-trough} duration as $\Phi_{X3} - \Phi_{X2}$. 
In the extreme case of high impact parameter, where the planetary disc only crosses the image of the stellar corona at grazing incidence angles, the maximum transit depth occurs at mid-transit ($\delta_{X,{\rm mid}} = \delta_{X,{\rm max}}$). In all the cases reported in this work, the planet crosses within the image of the stellar photosphere, making $\delta_{X,{\rm mid}} < \delta_{X,{\rm max}}$. 

The computations for transit maximum depth, average depth, duration, and total flux blocked can be applied similarly to the optical transit model. Table~\ref{tab:combRes} includes these values for both the X-ray and optical transit, using our fiducial model of the HD~189733 system.
%
The maximum X-ray transit depth is slightly shallower than that in the optical, and the average depth is significantly shallower. This is because the coronal X-ray emission is dispersed over a larger area overall, and the planet will only cross a small portion of the brightest emission ring at the edge of the stellar photosphere. Thus, most of the transit is spent crossing the dimmer X-ray regions, the coronal emission projected in front of the star. Additionally, the X-ray emission is spread out over a wider area of the sky, making the X-ray transit duration longer than the optical.
Nonetheless, the total signal (flux blocked) for the X-ray transit is comparable to that of the optical. 
Of course, this detectability estimator does not consider the orders of magnitude higher stellar flux that is present in optical observations, which makes actually detecting the transit far easier at those wavelengths. Thus, the collecting area of current and future X-ray telescopes remains the primary barrier for detecting X-ray transits.

These comparisons with the optical transit are highly X-ray model dependent. As we will demonstrate over the next few subsections, the X-ray model can change significantly depending on the assumed parameters. In each case, we provide values for the characteristics in Table \ref{tab:combRes}, so that they can be directly compared to the fiducial optical transit  model. Ideal models of a uniform stellar photosphere (e.g. finite size and the limb darkening effect being relatively small) make it so that the properties of an optical transit are easy to scale, with exceptions only in the case of very large planetary radii or high impact parameter. In the case of X-ray transits, the scaling properties are more subtle, and we explore them, below.

As a means of benchmarking the speed of the code, our fiducial case with 500 time bins takes $\sim 1$\,s to run. The calculation of the integration values for the coronal emission takes $\sim 0.25$\,s and the \texttt{batman} initialization takes much of the rest at just over 0.7\,s. Recalculating a new light curve from an already initialized model having changed parameters takes less than 0.1\,s. These values vary slightly depending on the model parameters, but each individual light curve takes no longer than a few seconds to run in all cases we tested, even if the number of time bins is increased to $>$10,000.

\subsection{Varying the absorber radius}

\begin{figure*}
    \centering
    \includegraphics[scale=0.60]{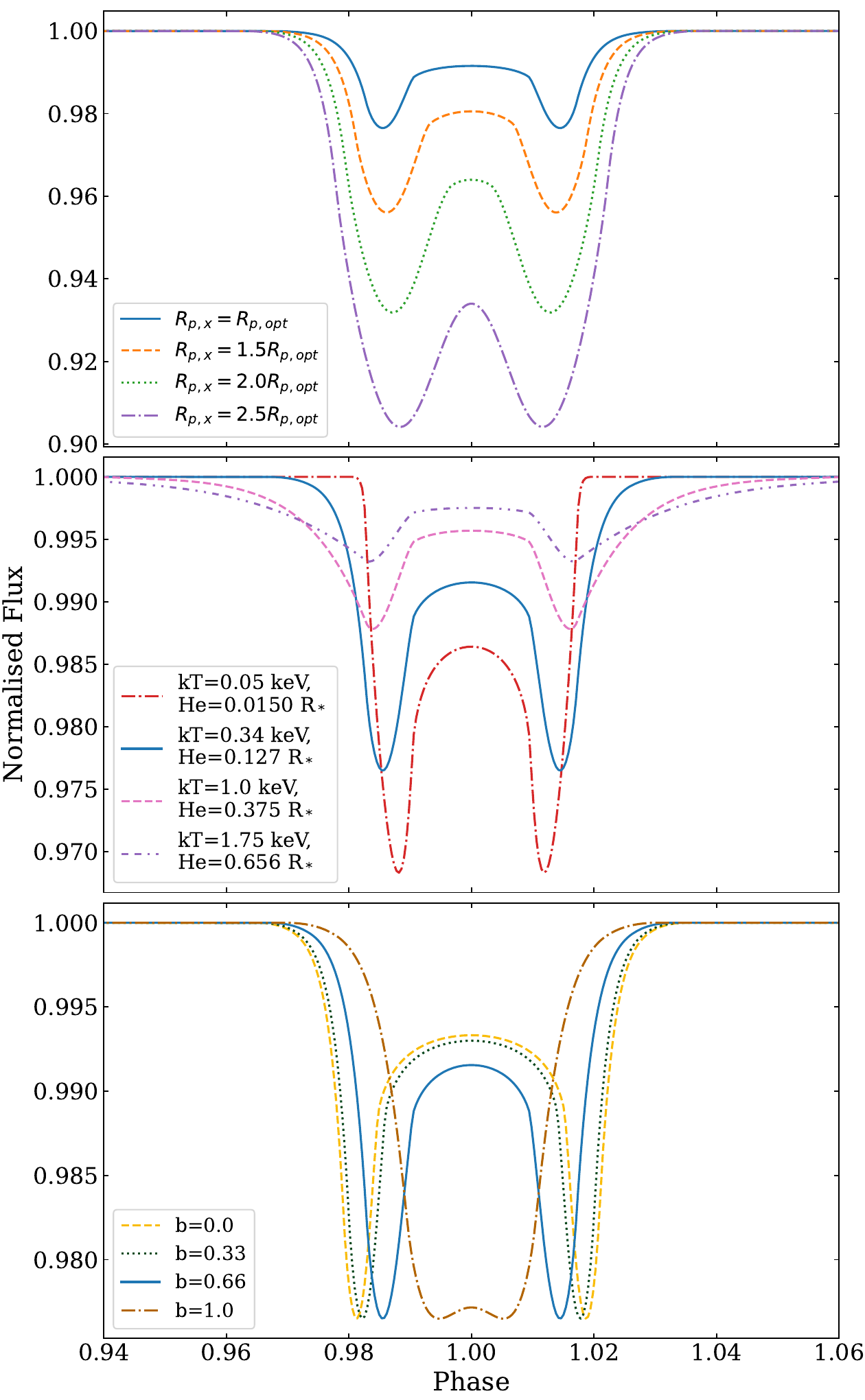}
    \caption{Comparison of transits of HD\,189733b resulting from varying different key parameters. In the top panel, we compare transits with four different absorber sizes for the same underlying coronal model. In the middle panel, we compare transits for four different coronal temperatures, assuming the effective X-ray radius of the planet is equal to the optical radius. In the bottom panel, we compare transits with four different impact parameters, assuming the same absorber size and coronal model in case. Calculated transit characteristics associated with the models in all three panels are given in Table \ref{tab:combRes}.}
    \label{fig:varyComb}
\end{figure*}

\begin{table*}
  \caption{Comparison of our model X-ray transit light curve properties. Our fiducial case the transit of HD\,189733b, assuming an absorber size equal to the optically measured radius, and a coronal temperature of 0.34\,keV. The other models then vary key parameters of interest to investigate their effect on the transit properties. We also include HD\,189733b's optical transit model parameters for comparison. The light curves associated with these properties are shown in Figs. \ref{fig:optR} (optical transit) and \ref{fig:varyComb} (variation of the X-ray model).}
  \label{tab:combRes}
\centering
\hspace*{-2.6cm}
\begin{tabular}{cccccccccc}
\hline
Absorber        & Coronal     & Emission     & Impact    & Maximum & Average & Mid-Transit & Transit  & Trough-to-      & Total X-ray    \\
Size            & Temperature & Scale Height, & Parameter, & Depth,               & Depth,         & Depth,               & Duration, & Trough Duration,  & Flux Blocked,   \\
$R_{\rm p,X}$   & $k_BT$      & $H_{\rm e}$   & $b$        & $\delta_{\rm X,max}$ & $\bar{\delta}$ & $\delta_{\rm X,mid}$ & $\Phi_D$  & $\Phi_{X3} - \Phi_{X2}$ & $\chi$          \\
(R$_{\rm opt}$) & (keV)       & ($R_*$)      &           & (\%)    & (\%)    & (\%)        & (Phase)  & (Phase)         & (\%)           \\ \hline

\multicolumn{10}{c}{\textit{Optical Model}} \\[0.1cm]

1.0             & n/a         & n/a          & 0.66      & 2.64    & 1.74    & 2.64        & 0.034    & n/a             & \textbf{0.0599} \\

\hline
\multicolumn{10}{c}{\textit{Fiducial HD\,189733 X-ray Model}} \\[0.1cm]

1.0             & 0.34        & 0.127        & 0.66      & 2.35    & 0.874   & 0.845       & 0.0653   & 0.0293          & \textbf{0.0574} \\

\hline
\multicolumn{10}{c}{\textit{Varying absorber size}} \\[0.1cm]

1.5             & 0.34        & 0.127        & 0.66      & 4.39    & 1.90    & 1.95        & 0.0677   & 0.0277          & \textbf{0.130} \\
2.0             & 0.34        & 0.127        & 0.66      & 6.81    & 3.25    & 3.60        & 0.0709   & 0.0261          & \textbf{0.233}  \\
2.5             & 0.34        & 0.127        & 0.66      & 9.59    & 5.15    & 6.60        & 0.0741   & 0.0236          & \textbf{0.369}  \\

\hline
\multicolumn{10}{c}{\textit{Varying coronal temperature}} \\[0.1cm]

1.0             & 0.05        & 0.0150       & 0.66      & 3.18    & 1.74    & 1.39        & 0.0381   & 0.0236          & \textbf{0.0667} \\
1.0             & 1.0         & 0.375        & 0.66      & 1.22    & 0.351   & 0.431       & 0.123    & 0.0325          & \textbf{0.0434} \\
1.0             & 1.75        & 0.656        & 0.66      & 0.678   & 0.174   & 0.248       & 0.193    & 0.0333          & \textbf{0.0336} \\ 

\hline
\multicolumn{10}{c}{\textit{Varying impact parameter}} \\[0.1cm]

1.0             & 0.34        & 0.127        & 0.0       & 2.35    & 0.778   & 0.668       & 0.0693   & 0.0373          & \textbf{0.0543} \\
1.0             & 0.34        & 0.127        & 0.33      & 2.35    & 0.805   & 0.700       & 0.0677   & 0.0357          & \textbf{0.0549} \\
1.0             & 0.34        & 0.127        & 1.0       & 2.35    & 1.04    & 2.28        & 0.0589   & 0.0108          & \textbf{0.0615} \\ \hline
\end{tabular}
\end{table*}

Here we examine how changing the effective radius of planetary X-ray absorption changes the transit model. An increase in the apparent radius of the planet could arise from strong X-ray absorption by an extended or escaping atmosphere. The top panel of Fig. \ref{fig:varyComb} demonstrates how the X-ray transit depth changes if we set $R_{\rm p,X}/R_*$ to a value that is 1.5, 2, and 2.5 times the optically measured radius. 
Table~\ref{tab:combRes} reports the corresponding benchmark parameters for the light curve shape and signal. 
For limb darkened photospheric models in the visible and near-infrared, the transit depths and signal usually scale with $\left(\frac{R_{\rm p,opt}}{R_*}\right)^2$, because variation of the emission intensity across the stellar disc is small. The values in Table~\ref{tab:combRes} demonstrate that this is not always the case for X-ray transits, due to the more complex distribution of the coronal flux. For example, the maximum depth values exhibit large deviations from a quadratic scaling rule. Other depth parameters show much smaller deviation. We explore this further in Section~\ref{sec:scalings}.

The morphology of the transit profile changes somewhat between the 1.0 and 2.5\,R$_{\rm p,opt}$ cases. This can be seen visually in Fig. \ref{fig:varyComb}, but also in the calculated quantities in Table \ref{tab:combRes}. As the absorber radius increases, the ``mid-transit depth" gets much deeper. In the 2.5\,R$_{\rm opt}$ case, this point is nearly 8 times deeper than for the 1.0\,R$_{\rm opt}$ transit. This occurs because the relatively high impact parameter of \hdb\ leads to some part of the X-ray absorbing region always blocking an arc of the bright disc edge, even at mid-transit.

For the larger absorbers, the point at which they cover the largest arc of the bright ring around the stellar disk is slightly closer to the mid transit point, and so the trough-to-trough duration of the transit reduces. 
This also makes it so that the duration of the flat mid-transit region shrinks for larger absorber sizes. For the 2.5\,R$_{\rm opt}$ transit, this is reduced to a single minimum flux point. We finally note that the individual troughs themselves are wider for the larger absorber sizes, and the full transit duration is slightly longer.

\subsection{Varying the coronal scale height}
\label{ssec:HVariation}

\begin{figure*}
    \centering
    \includegraphics[scale=0.64]{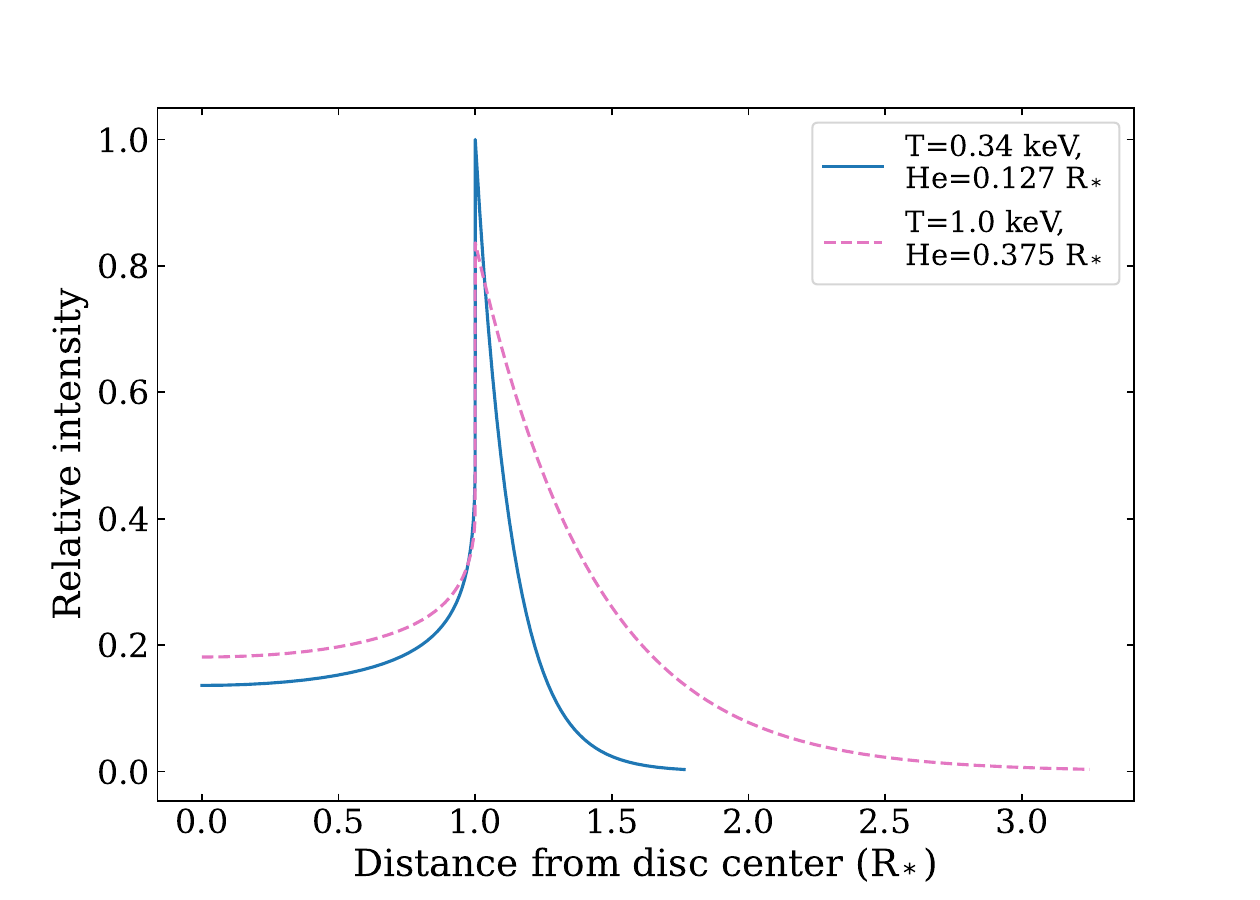}
    \includegraphics[trim={2cm 0 1cm 2cm},clip,scale=0.36]{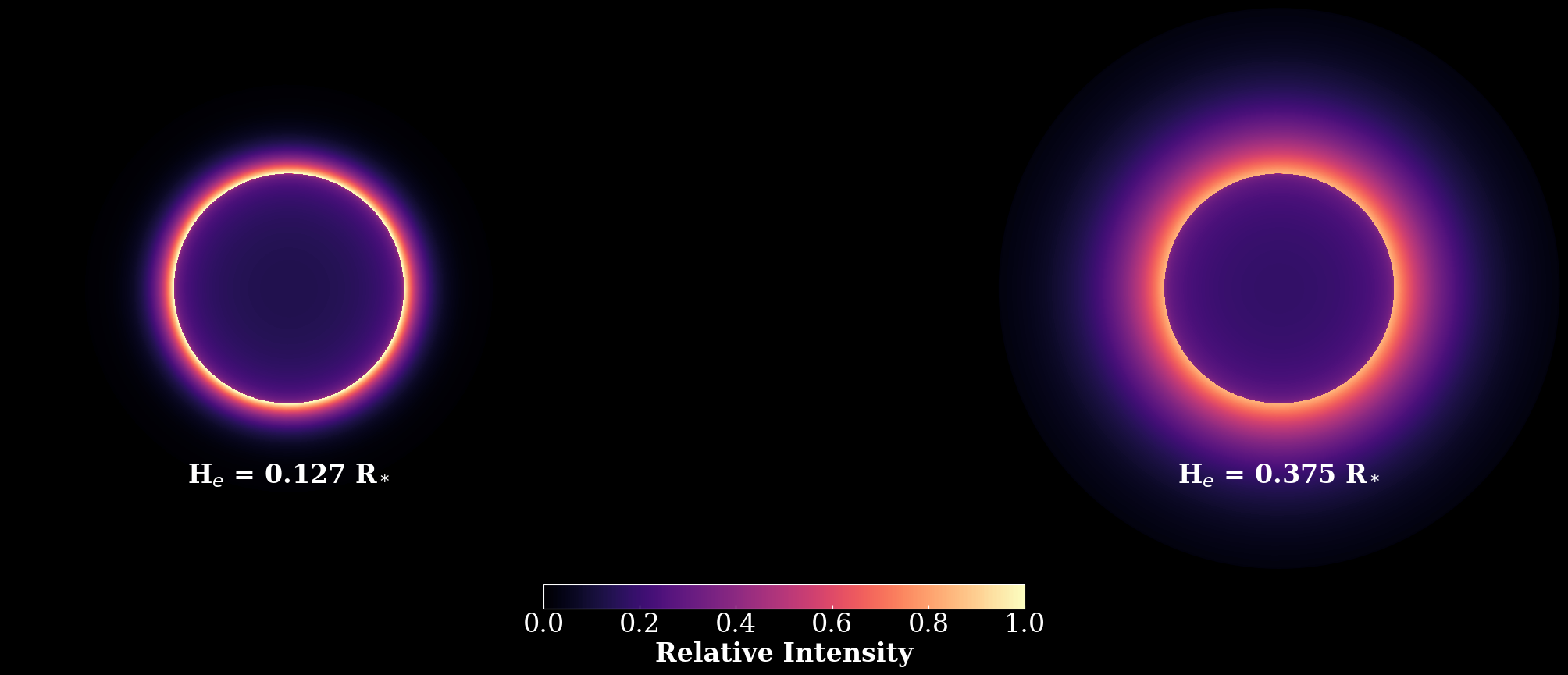}
    \caption{The effect of changing the coronal scale height/temperature on the emission profile. In the top panel we show the relative 1-D intensities as a function of distance from the center of the stellar disk associated with the $H_{\rm e} = 0.127$ and 0.375\,R$_*$ ($kT = 0.34$ and 1.0\,keV) models. In the bottom panel we show the 2D representations of the stellar X-ray emission profiles associated with the same models. The relative emissions are plotted on the same scale, which is $I(x)$ renormalized to the maximum coronal emission intensity of the $H_{\rm e} = 0.127$\,R$_*$ model. Transit profiles associated with these models are plotted in the middle panel of Fig. \ref{fig:varyComb}.}
    \label{fig:ChangeH}
\end{figure*}

The greater complexity of the coronal emission distribution compared to optical limb-darkening results in a wider range of possible light curve morphologies. To explore this further, we next varied the coronal model itself by changing the coronal temperature, and thus the emission scale height. We use three alternative coronal temperatures: 0.05, 1.0, and 1.75\,keV, which for HD\,189733 correspond to emission scale heights of 0.0150, 0.375, and 0.656\,$R_*$.

In Fig. \ref{fig:ChangeH}, we compare the emission profiles in 1-D and 2-D for two of our models: the original 0.34\,keV temperature, and the 1.0\,keV temperature. For higher temperatures, the emission is spread out over a much larger area. Simultaneously, this \textit{reduces} the maximum intensity of the emission at the edge of the photospheric disc, as compared to a lower temperature model.

Consequently, transits over less compact coronae are shallower, but have a longer duration. To demonstrate this, we show transit profiles for the four coronal models we considered in the middle panel of Fig. \ref{fig:varyComb}, while key quantitative characteristics of the transit profiles are given in Table \ref{tab:combRes}. The table demonstrates that overall detectability decreases for transits across higher temperature coronae. While the increased durations act to partially offset the reduced transit depths, there is still 24\% less flux blocked in the 1.0\,keV case compared to the 0.34\,keV case. This makes sense, as for a fixed absorber size, increasing $H_{\rm e}$ spreads the light across a wider area and so the total light intercepted across the transit accordingly decreases. 

We note that as the coronal temperature is reduced and the plasma becomes more compact, the maximum depth appears to approach the approximation for a geometrically thin emission presented in equation 2 of \citet{Schlawin2010}. For HD\,189733b, the maximum depth predicted by their relation is 3.09\%, as compared to the 3.18\% for the most compact corona we tested, assuming $R_{\rm p,X} = R_{\rm opt}$. While these values are compatible, X-ray emission is expected to be extended beyond the photospheric disk, and so the usefulness of this approximation in the general case in the coronal X-ray regime is unclear. We emphasize that our work intrinsically handles this extended emission, and also does not use the same geometrical approximations for the shape and size of the shadow which limit the precision of and regime in which the \citet{Schlawin2010} equation is valid.

As the coronal temperature increases, the points of maximum transit $\Phi_{X2}$ and $\Phi_{X3}$ shift further away from the transit center ($\Phi_C$), leading to larger trough-to-trough durations in Table \ref{tab:combRes}. This results from the increased scale height leading to a slower drop off in the emission, such that the point of maximal emission coverage is slightly further from the edge of the stellar photosphere. Looking at the top panel of Fig \ref{fig:ChangeH}, a \hdb-like absorber with radius $R_{\rm p,opt} \approx 0.3~R_*$ will cover the maximum integrated flux from the 0.34~keV curve later in the transit as compared to the 1~keV curve, where the coronal emission is more spread out.

\subsection{Varying the impact parameter}

Of the three model variables investigated in this work, the impact parameter, $b$, is usually the best constrained from optical/near-IR wavelength discovery. 
We now test how the transit light curve changes with the impact parameter, where $b=1$ refers to the usual case of grazing incidence with the photospheric disk, and not $x=1$. 
We test three values in addition to the fiducial case: $b=$ 0, 0.33, and 1. The bottom panel of Fig. \ref{fig:varyComb} compares the transits for these four different impact parameters. All other parameters were held at their HD\,189733 system values.

The effects of increasing the impact parameter appear relatively small until $b$ approaches 1. The bottom panel of Fig. \ref{fig:varyComb} demonstrates that impact parameter has a negligible effect on the maximum transit depth. Instead, as is the case for optical transits, impact parameter significantly affects the transit duration, with larger impact parameters resulting in shorter transits.
As $b$ approaches 1, the path of the planet across the projected coronal emission begins to intersect with the arc of highest emission for more of the transit, leading to larger $\delta_{X,{\rm mid}}$. This means that in addition to the overall duration shortening, the maximum transit times $\Phi_{X2}$ and $\Phi_{X3}$ shift closer to the transit center, reducing the trough-to-trough duration. In the extreme case of $b=1$, the flux rise from $\Phi_{X2}$ to $\Phi_C$ almost entirely disappears, and the transit morphology instead resembles that of an optical transit with a spot crossing event. Even though the transit duration $\Phi_D$ gets smaller with higher impact parameter, the average depth $\bar{\delta}$ is larger, making $b \approx 1$ transits slightly more detectable as gauged by the total flux blocked, $\chi$ (Table~\ref{tab:combRes}). 

We also note that X-ray observations could detect transits for $b > 1$. Some planets with $b$ close to but above unity can be detected via optical transits \citep[e.g., HIP~65~A~b][]{Nielsen2020}. In principle, X-ray observations could increase the number of planet transits detected at high impact parameter, where there may be no overlap of the planet and stellar discs in the optical. Both the extended nature of the coronal emission beyond the photospheric disc of the star, as well as the possibility that the X-ray absorbing region is larger than the optical, makes it easier to detect $b \geq 1$ transits. A possible example of this in Ly-$\alpha$ was described by \citet{Ehrenreich2012} for 55\,Cnc\,b, where they attribute a drop in flux around the expected time of inferior conjunction to an extended atmosphere of the planet, despite it being non-transiting in the optical.

\subsection{Effects of light curve binning}
\label{sec:Binning}

\begin{figure*}
    \centering
        \includegraphics[scale=0.64]{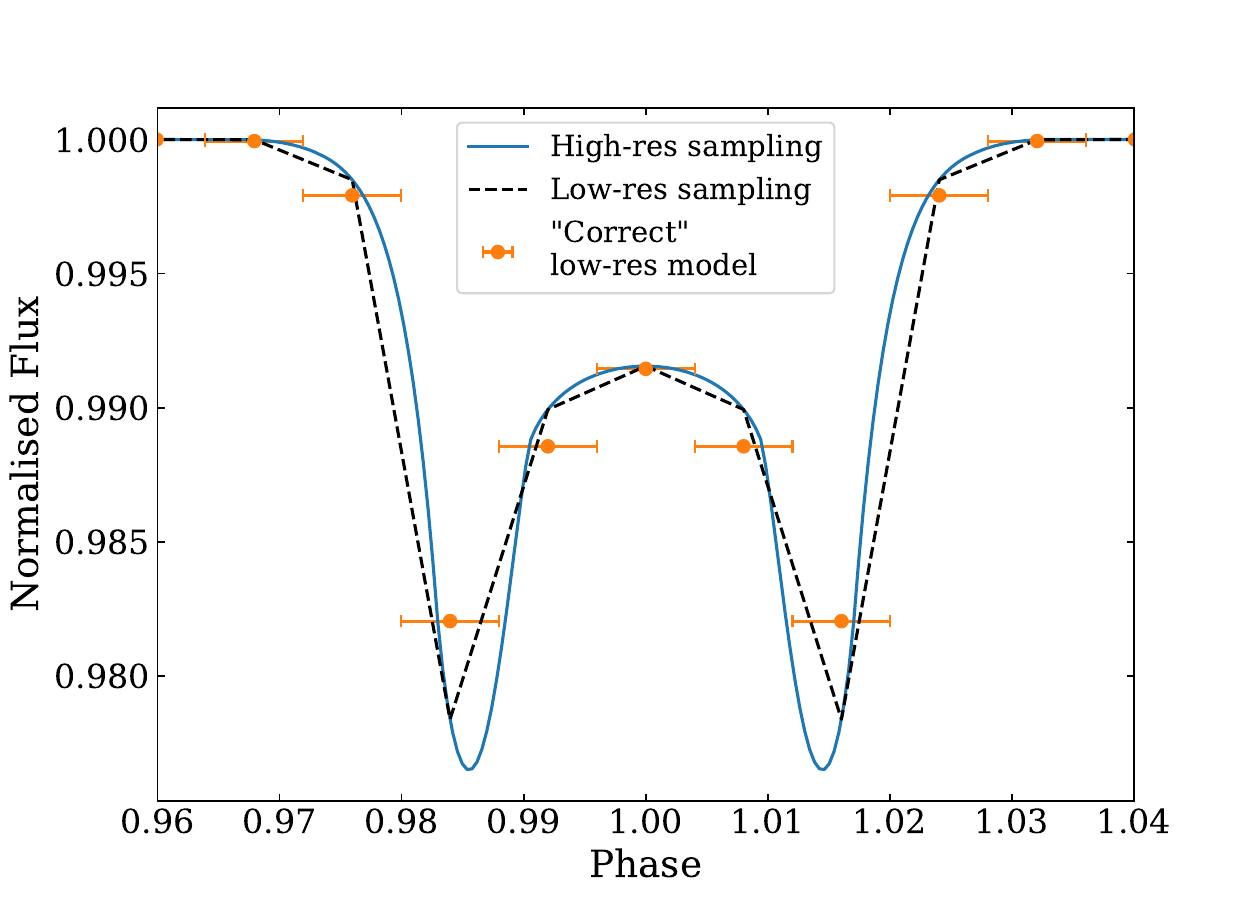}
    \caption{The effect of changing the time resolution with the same underlying model. The ``high-res sampling" model is the original 500 bin model, also plotted in e.g. Fig. \ref{fig:optR}. The ``correct low-res model" takes the original 500 bin model and averages across groups of $\sim$20 points to obtain a representative value of the model for the phase range each resulting point covers. The 25 phases between 0.9 and 1.1 that these points are plotted at were then ran through \batman to obtain an instantaneous value of the model at those phases for comparison - these are plotted as the ``low-res sampling". This demonstrates why the X-ray transit model we present should always be calculated at high resolution first, and then binned to the cadence of the data it is being compared to.}
    \label{fig:timeRes}
\end{figure*}

We also examined the effect of time resolution on the output model. To compare to our original model with 500 bins (equivalent to 76.7\,s resolution for \hdb), we generated two other model light curves both with 25 bins between 0.9 and 1.1 in phase, plotted in Fig. \ref{fig:timeRes}.
The first of these was obtained by averaging sets of $\sim$20 points from the original 500 bin model, in order to obtain 25 roughly evenly-spaced bins at lower phase resolution (orange points). Each set of $\sim$20 points cover a non-negligible phase range, with the averaged value being representative of that phase range. To compare with this, for the second we took the 25 phases of that model and ran them directly through batman (black dashed). In simply providing the instantaneous value of the model at the plotted phases, these are not necessarily representative of the model across all phases surrounding each point.

The two show some marked deviations from one another. The black dashed model aligns almost exactly with the higher resolution 500 bin model at every point for which it was calculated. However, since the binned model represents a range of phases, it departs from the other two plotted models. This effect is largest in regions where the derivative of the underlying model is changing appreciably, such as close to the phases of maximum transit depth. For real data, due to the very low count rates typically measured at X-ray wavelengths, data will likely require binning to a lower resolution in this way. Such time bins are best fit by an average of the underlying model sampled through the time range covered, as opposed to an instantaneous evaluation of the model at the plotted bin center. When assessing observational data, one should always calculate these models at higher resolution and bin to the lower resolution of the actual data.

\section{Investigating scaling relations}
\label{sec:scalings}

\begin{figure*}
    \centering
    \includegraphics[width=\textwidth]{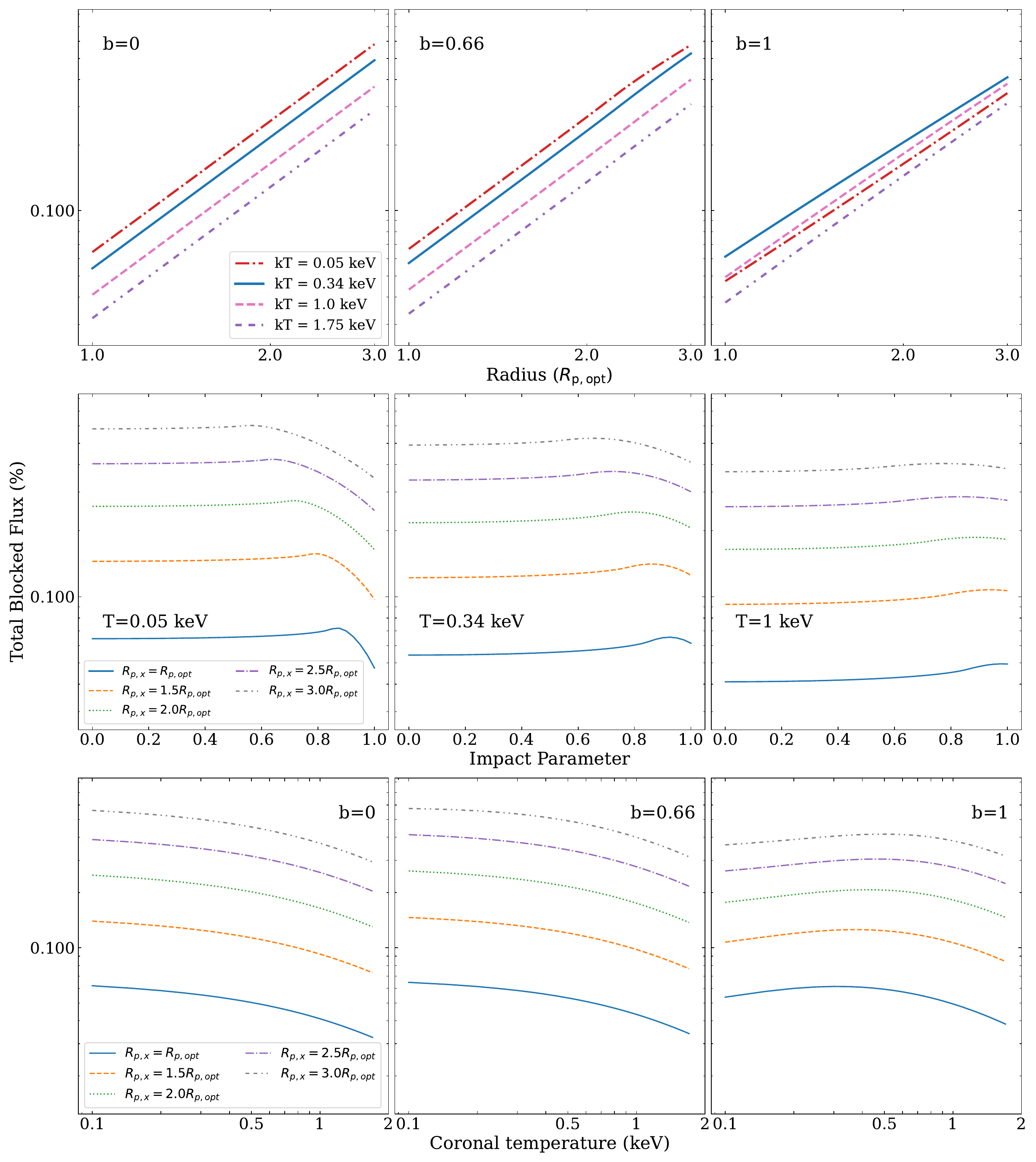}
    \caption{Change in the total flux blocked ($\chi$, a proxy for transit detectability) while varying three different parameters. The top panels show the change in $\chi$ with increasing absorbing region radius at fixed values of the impact parameter and coronal temperature. The middle panels show the change in $\chi$ with increasing impact parameter at fixed values of the absorber radius and coronal temperature. The bottom panels show the change in $\chi$ with increasing coronal temperature at fixed values of the absorber radius and impact parameter. Each panel's label shows the quantity that is fixed across all results in that panel. Each legend shows the values of the other fixed quantity on that row, where each line style and color combination on that row refers to a different fixed value of that parameter.}
    \label{fig:detect_varyAll}
\end{figure*}

As demonstrated above, the observed shape and depth of an X-ray transit is strongly impacted by stellar coronal temperature and the impact parameter for the planetary orbit. These factors can affect the overall detectability of a transit, because X-ray detector count rates are generally low for most stars and thus require broad time bins that might wash out deeper features in the transit light curve (Section~\ref{sec:Binning}). Here we examine how one comprehensive metric -- the total flux blocked by the transit ($\chi$, Equation~\ref{eq:FluxBlocked}) -- might scale with parameters investigated in this work, providing a general rule-of-thumb by which one can gauge detection feasibility.

The top panels of Fig.~\ref{fig:detect_varyAll} shows the results of changing the apparent X-ray radius of the planet, for multiple (but fixed) values of the coronal temperature and impact parameter. 
By examining the trends for $\chi$ in log-log space, we can see that there is a quadratic 
increase in the total light blocked with increasing planetary radius, for almost all cases with $b \leq 0.66$. 
This is reminiscent of optical transits where the maximum depth of the transit is proportional to the geometric projection of the planetary disk, $\left(R_{\rm p,opt}/R_*\right)^2$. 
This relationship only breaks down when the impact parameter and/or radius of the planet is large enough such that only a portion of the planetary disk blocks the photosphere at transit conjunction. Similarly, in the case of X-ray transits, when the impact parameter and/or effective radius is large enough to extend beyond several coronal emission scale heights, the quadratic scaling relation breaks down. This is most apparent for the lowest temperature ($kT = 0.05$~keV) model, which has the smallest coronal scale height.

For the grazing incidence, $b=1$ case, the best fit power laws range from 1.72 to 1.92. Notably, the total X-ray flux blocked in the $kT = 0.05$ is smaller across all planetary radii than the warmer $kT = 0.34$ and $1$~keV models, contrary to the results at other impact parameters. This is likely because the planet radius extends beyond the limits of our coronal surface brightness model ($6 H_{\rm e}$ above the photosphere). We encourage researchers to adjust our coronal model to evaluate limiting cases such as this one, if needed.

The middle row of Fig.~\ref{fig:detect_varyAll} shows the results of varying the transit impact parameter, while holding the absorber radius and coronal temperature fixed. The curves remain very close to flat until the impact parameter is large enough for the planetary disc to block the stellar corona at grazing incidence angles. The curves also demonstrate how this threshold changes depending on radius and coronal temperature. Cooler temperature coronae are less extended, so the threshold $b$ for more complex scaling gets smaller with larger planetary radii (left panel of the middle row of Fig.~\ref{fig:detect_varyAll}). When the planetary radius is so large that it blocks several scale heights of the corona and less of the photospheric disk, the transit blocks less X-ray flux overall. However, in some cases, this grazing incidence causes the transit to block more light than the $b=0$ case because the path of the planetary disc covers more of the limb-brightened portion of the corona, as is the case for smaller planetary radii in the $T=0.34$ and $1$\,keV subplots in the middle row of Fig.~\ref{fig:detect_varyAll}. 

The bottom panels of Fig.~\ref{fig:detect_varyAll} shows the results of varying the coronal temperature, holding the absorber radius and impact parameter fixed. In this case, the trends are less simple, with no single power law holding as the temperature is varied. 
Nonetheless, the curves appear parallel to each other, their curvature is small, and take on the same shape for impact parameters $b \leq 0.66$, while the relationships break down for grazing incidence angles. On average, $\chi$ scales roughly with a power law slope of -1/4. The actual change in $\chi$ will be less than this at lower coronal temperatures, and greater at higher temperatures.
With all the information provided above, we are able to produce an empirical, approximate relationship for the overall strength of the X-ray transit via the amount of light blocked ($\chi$)
\begin{equation}
    \chi \approx 5.4 \times 10^{-4}\  
        \left(\frac{R_{\rm p,X}/R_*}{0.156}\right)^{2} 
        \left(\frac{kT}{0.34~{\rm keV}}\right)^{-1/4}
\end{equation}
for $b \leq 0.66$ and $R_{\rm p,X}/R_* \leq 0.4$, based on the fiducial parameters of the \hdb\ system (Section~\ref{sec:Fiducial}), but with $b=0$. This expression differs from equation 2 of \citet{Schlawin2010} by providing a way of estimating the overall transit detectability, as opposed to the maximum transit depth. As shown in Table~\ref{tab:combRes}, while maximum depth can be a reasonable proxy for detectability, transits with a larger maximum depth do not always have a larger value of $\chi$.

\section{Limitations of the model}

There are a few of limitations to calculating X-ray transit light curve models in the way we have introduced in this paper. We discuss them here, together with some suggestions for how the effect of these may be mitigated. Example cases for implementing various mitigation strategies are included in the public release of the code.

\subsection{Isothermal corona}

\begin{figure*}
    \centering
        \includegraphics[scale=0.56]{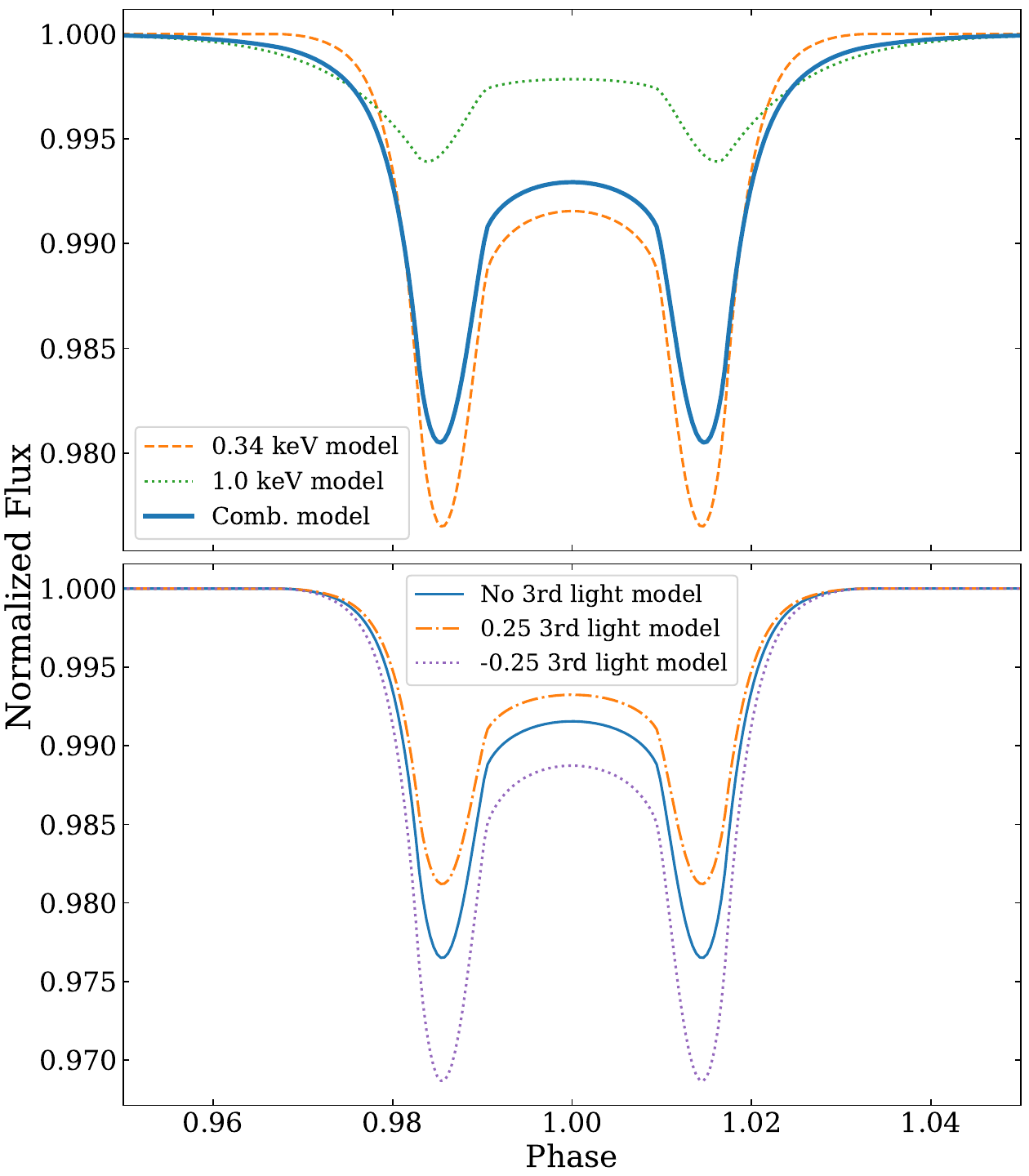}
    \caption{Examples of methods for mitigating the limitations of our model. In the top panel, we consider a possible mitigation of the isothermal nature of our coronal model by combining the model arrays of two separate models. Here, we use the same 0.34\,keV and 1.0\,keV models we used in Section \ref{ssec:HVariation}. We combine their model fluxes with a factor 0.5 in the 1.0\,keV to reflect a case where the count rate of photons from the plasma at that temperature is half that from the lower temperature plasma. In the bottom panel, we give examples of including a third light parameter in the model, in order to partly mitigate the symmetric nature of our model. The positive and negative third light reflect transit chords across latitudes that are dimmer and brighter, respectively, as compared to the rest of the corona.}
    \label{fig:mitigation}
\end{figure*}

Firstly, our model assumes an isothermal corona, which is implemented in the model as a single scale height value. In reality, we know from measuring the X-ray spectra of stars that emission occurs over a wide, continuous range of temperatures \citep[for a discussion, see e.g.][]{Guedel2004}. In section \ref{ssec:HVariation}, we demonstrated the effect of varying the single scale height on the obtained transit light curve, which led to significant changes in the transit depth and shape.

To better simulate the complexities of coronal structure, one could combine multiple models with different temperatures, and therefore scale heights, additively. This would be in a similar vein to how stellar X-ray spectra are often fitted, where models representing a few characteristic temperatures are combined to produce the final model. The individual models could further be scaled according to some normalisation such that they produce different magnitudes of relative contribution to the final model.

The top panel of Fig. \ref{fig:mitigation} displays an example of combining two scale heights, using the same two temperatures we used in Section \ref{ssec:HVariation}. We combined them after having calculated the light curve model for each separately, and included a factor of 0.5 in the 1\,keV model to set the level of its total integrated emission compared to the 0.34\,keV model. The precision of current and immediate future data will likely not be sufficient to warrant models with more than a couple of temperatures due to the small magnitude of changes to the transit morphology we observe in these combined models.

\subsection{Symmetric corona}

Our model assumes the corona is radially symmetric in 3-D. In optical light, aside from a few small regions associated with starspots, variations are typically gradual across the disc (e.g. that due to the limb darkening effect). In contrast, a substantial portion of stellar X-ray emission arises from magnetically active regions, and is therefore inherently inhomogeneous. The effect of this activity on the morophology of an individual transit event can be substantial, as demonstrated by \citet[see e.g. their fig. 2]{Llama2015}.

However, with the current generation of X-ray telescopes and their likely immediate successors, detection of X-ray transits in single or even a few observations is highly unlikely for all known systems \citep{Cilley2024}. This is due to the low count rates one attains for even the brightest targets with large transiting planets. For the foreseeable future, any detections with a good enough precision to warrant detailed modelling will very likely require combining at least 10-20 observations together. This will average over the inhomogeneities, reducing the importance of individual active regions.

There are some issues that averaging will not be able to help with. For instance, a star could preferentially emit X-rays from particular (ranges of) latitudes. One could incorporate this into the model by adding in a ``third light", $F_{\rm 3L}$, parameter such that
\begin{equation}
F_{\rm final} = \frac{F_{\rm no3L} + F_{\rm 3L}}{1 + F_{\rm 3L}},
\label{eq:thirdL}
\end{equation}
where $F_{\rm no3L}$ is the flux from the usual model not including the third light, and $F_{\rm final}$ is the final, adjusted flux. ``Third light" parameters are typically included in studies of eclipsing binaries where an unseen third star is known or suspected to contribute to the light curve. The effect of a third light here would account for the transit occurring on a particularly dim chord by considering the non-transit areas of the disc brighter. In the case of a transit crossing a brighter than average chord, $F_{\rm 3L}$ would take a negative value. We also note that Equation \ref{eq:thirdL} still assumes that the model underlying the transit chord is symmetric. If required, more complex modeling of transits across recurrent, bright, and localized active regions could be handled by a code such as \texttt{starry} \citep{Luger2019}, and combined with our model of the transit across the quiescent emission.
In the bottom panel of Fig. \ref{fig:mitigation} we show examples of including positive and negative third light parameters in the model. Using the method in Equation \ref{eq:thirdL} works to suppress or increase the strength of the dips, without altering the overall shape of the transit.

\subsection{Spherically symmetric, opaque absorber}

Our model also assumes the absorbing region is fully opaque and circular in shape. While this is usually true across the optical disk of the planet, in the upper atmosphere of the planet some fraction of the X-ray light may pass through the atmosphere unaffected. Moreover, the absorber need not be circular, particularly if atmospheric escape leaves behind co-orbital material. For example, the Ly-$\alpha$ transit of GJ\,436\,b is very asymmetric, with a large comet-like tail of absorbing hydrogen gas stretching out for many planetary radii behind the planet itself \citep{Ehrenreich2015,Lavie2017}. The transparency level and non-circular morphology would also likely be somewhat degenerate in the model, and one may struggle to differentiate between a circular absorber whose transmission is greater in some regions planet's upper atmosphere, and a highly non-circular absorber that could be fully opaque or translucent in some areas.

\section{Conclusions}

We have adapted the popular exoplanet transit code \textsc{batman} to generate models of X-ray transits. Our idealized stellar corona is isothermal and optically thin, with radially symmetric emission in 3 dimensions. While this ignores the effect of inhomogeneity from e.g. active regions, detecting transits with current and near-future X-ray telescopes will require at least several epochs of observations to be combined, which will smear out the effect individual bright regions.

We anticipate that our adaptations will be useful for observers and modellers alike, allowing the relatively fast calculation of thousands of X-ray transit profiles. Our code is publicly available, together with examples of its use, our search for scaling relations, and methods for mitigating some limitations. These examples include the effect of varying the size of the X-ray-absorbing region and the impact parameter of the planet, as well as the coronal temperature. Together, varying these parameters produces a wider array of typical transit morphologies than typically observed at optical wavelengths.

We empirically derived scaling relationships for how our defined measure of the overall transit detectability, $\chi$, scales when varying the same three parameters. We found that $\chi \propto R_{\rm p,X}^2$, across much of the tested parameter space. This is analogous to how optical and near-IR transit depths scale with the square of the planet radius, even though the maximum X-ray transit depth does not seem to scale simply with the absorber size. When we varied the coronal temperature, the scaling with $\chi$ was less simple, but averages out to a power law relationship of about $-1/4$ for $b \lesssim 0.66$. Varying the impact parameter has little effect on detectability until the planet approaches $b=1$. 

We expand on the work presented here in \citet{Cilley2024}, in which we apply our transit model to a large sample of known exoplanets, and identify the top targets for future observation.

\begin{acknowledgments}
We thank the anonymous referee for their insightful report which helped improve this manuscript. We also thank Laura Kreidberg for providing the community with \textsc{batman}, as well as her ongoing support of this excellent package.
\end{acknowledgments}

%

\vspace{5mm}


\software{astropy \citep{Astropy},
          batman \citep{batman},
          matplotlib \citep{matplotlib,matplotlib334},
          scipy \citep{scipy}
          }






\bibliography{main}{}
\bibliographystyle{aasjournal}



\end{document}